\documentclass[a4paper,reprint]{revtex4-1}

\usepackage{graphicx}
\usepackage{color}
\usepackage{placeins}

\usepackage{amsmath}
\usepackage{mathtools}

\usepackage{siunitx}
\DeclareSIUnit\Molar{\textsc{M}}

\begin{document}
\title{Wedge wetting by electrolyte solutions}

\date{\today}

\author{Maximilian Mu{\ss}otter}
\email{mussotter@is.mpg.de}
\author{Markus Bier}
\email{bier@is.mpg.de}
\affiliation{
	Max Planck Institute for Intelligent Systems, 
	Heisenbergstr.\ 3,
	70569 Stuttgart,
	Germany
}
\affiliation{
	Institute for Theoretical Physics IV, 
	University of Stuttgart,
	Pfaffenwaldring 57,
	70569 Stuttgart,
	Germany
}
\begin{abstract}
	The wetting of a charged wedge-like wall by an electrolyte solution is 
	investigated by means of classical density functional theory.
	As in other studies on wedge wetting, this geometry is considered as the most
	simple deviation from a planar substrate, and it serves as a first step towards
	more complex confinements of fluids.
	By focusing on fluids containing ions and surface charges, features of real
	systems are covered which are not accessible within the vast majority of 
	previous theoretical studies concentrating on simple fluids in contact with 
	uncharged wedges.
	In particular, the filling transition of charged wedges is necessarily of first
	order, because wetting transitions of charged substrates are of first order and
	the barrier in the effective interface potential persists below the wetting
	transition of a planar wall; hence, critical filling transitions are not expected 
	to occur for ionic systems. 
	The dependence of the critical opening angle on the surface charge, as well as
	the dependence of the filling height, of the wedge adsorption, and of the line
	tension on the opening angle and on the surface charge are analyzed in detail.
\end{abstract}
\maketitle
\section{Introduction}

Over the past few decades numerous theoretical and experimental investigations
have been performed aiming at a microscopic understanding of the phenomena of 
fluids at interfaces, e.g., capillarity, wetting, and spreading, which are of
technological importance for, e.g., coating processes, surface patterning,
or the functioning of microfluidic devices \cite{Lin2011,Vogel2012,Nee2015,
Wen2017,GalindoRosales2018}. 
Particularly simple model systems to investigate these phenomena 
theoretically are planar homogeneous substrates, which have been studied
intensively \cite{Dietrich1988,Schick1990,Dietrich1991}.
This way, methods have been developed to relate the thickness of fluid films
adsorbed at substrates and the contact angle to fluid-fluid and wall-fluid
interactions, to infer surface phase diagrams, and to characterize the order
of wetting transitions.

However, the preparation of truly flat homogeneous substrates requires a huge
technical effort and in nature there is no such thing as a perfectly flat surface 
\cite{Lueth2015}.
On the one hand, one is always confronted with geometrically or chemically 
structured substrates, irregularly-shaped boundaries, or geometrical disorder. 
On the other hand, modern surface patterning techniques allow for the 
targeted fabrication of structured substrates with pits, posts, grooves, 
edges, wedges etc.\ in order to generate functionality, e.g., superhydrophobic
surfaces \cite{Checco2014}. 
This leads to the necessity of studying substrates beyond the simple flat 
geometry, but the wetting properties of such nonplanar substrates are very 
different from smooth and planar walls and their description is much more 
complex.

Perhaps the most simple of the aforementioned elementary topographic 
surface structures are wedges, which are formed by the intersection of two
planar walls meeting at a particular opening angle.
First predictions of the phenomenon of the filling of a wedge upon decreasing
the opening angle have been based on macroscopic considerations 
\cite{pomeau1986wetting, hauge1992macroscopic}.
Microscopic classical density functional theory and mesoscopic approaches based
on effective interface Hamiltonians revealed that systems with long-ranged
Van-der-Waals interactions, where critical wetting transitions of planar walls
occur, exhibit critical wedge filling transitions with universal asymptotic
scaling behavior of the relevant quantities \cite{NAPIORKOWSKI1992,
napiorkowski1994wetting, parry1999universality}.
It has been argued that the order of a filling transition equals the order of
the wetting transition of a planar wall \cite{Rejmer1999}.
However, it turned out later that the relation between the orders of 
wetting and filling transitions is more subtle: If the wetting transition
is critical then the filling transition is critical, too. Otherwise, if the 
wetting transition is of first order then the filling transition may be 
first-order or critical, depending on whether or not a barrier exists in the
effective interface potential at the filling transition 
\cite{parry2000critical, parry2001interfacial}.
A consequence of the latter scenario with first-order wetting transitions is
the possibility to have first-order filling transitions, if the critical opening
angle is wide, and critical filling transitions, if it is narrow.
These predictions from mesoscopic approaches have been recently verified by
microscopic classical density functional theory \cite{malijevsky2013critical,
malijevsky2015filling}.

In order to reduce complexity, all cited previous theoretical studies on
wedge wetting have been performed for models of simple fluids.
However, many fluids used in applications, including pure water due to its 
autodissociation reaction, are complex fluids containing ions, so that the
generic situation of wedge wetting by electrolyte solutions is of enormous 
interest from both the fundamental as well as the applied point of view.
Despite the huge relevance of electrolytes as fluids involved in wedge wetting
scenarios \cite{baratian2015shape}, this setup has not been theoretically 
studied before on the microscopic level, probably due to the expected lack of
universality and increased complexity as compared to cases with critical
wetting and filling transitions. 
Indeed, it turned out for planar walls that the presence of ions, not too close to bulk
critical points, generates first-order wetting and a non-vanishing barrier in
the effective interface potential below the wetting transition 
\cite{Ibagon2014}.
Hence, on very general grounds, one expects first-order filling transitions of
wedges to take place for electrolyte solutions.

In the present work, a microscopic lattice model is studied within a classical
density functional theory framework in order to investigate the properties of
wedge wetting by electrolyte solutions. The usage of a lattice model allows
for technical advantages over continuum models \cite{Ibagon2013,Ibagon2014,Ibagon2016}.
The model and the density functional formulation is specified in Sec.~\ref{sec:theory}.
In Sec.~\ref{sec:results} first the bulk phase diagram and the wetting behavior of a
planar wall of the considered model are reported.
Next, wedge wetting is studied in terms of three observables: the wedge
adsorption, the filling height, and the line tension.
The dependence of these quantities on the wedge opening angle, on the surface
charge density of the walls of the wedge, as well as on the strength and the range of the 
nonelectrostatic wall-fluid interaction are discussed in detail.
Concluding remarks on the first-order filling transition considered in
the present work and the more widely studied critical filling transition are
given in Sec.~ \ref{sec:conclusion}. 

\section{Theoretical foundations\label{sec:theory}}

\subsection{Setup}
\begin{figure}[t]
	\includegraphics[width=8cm]{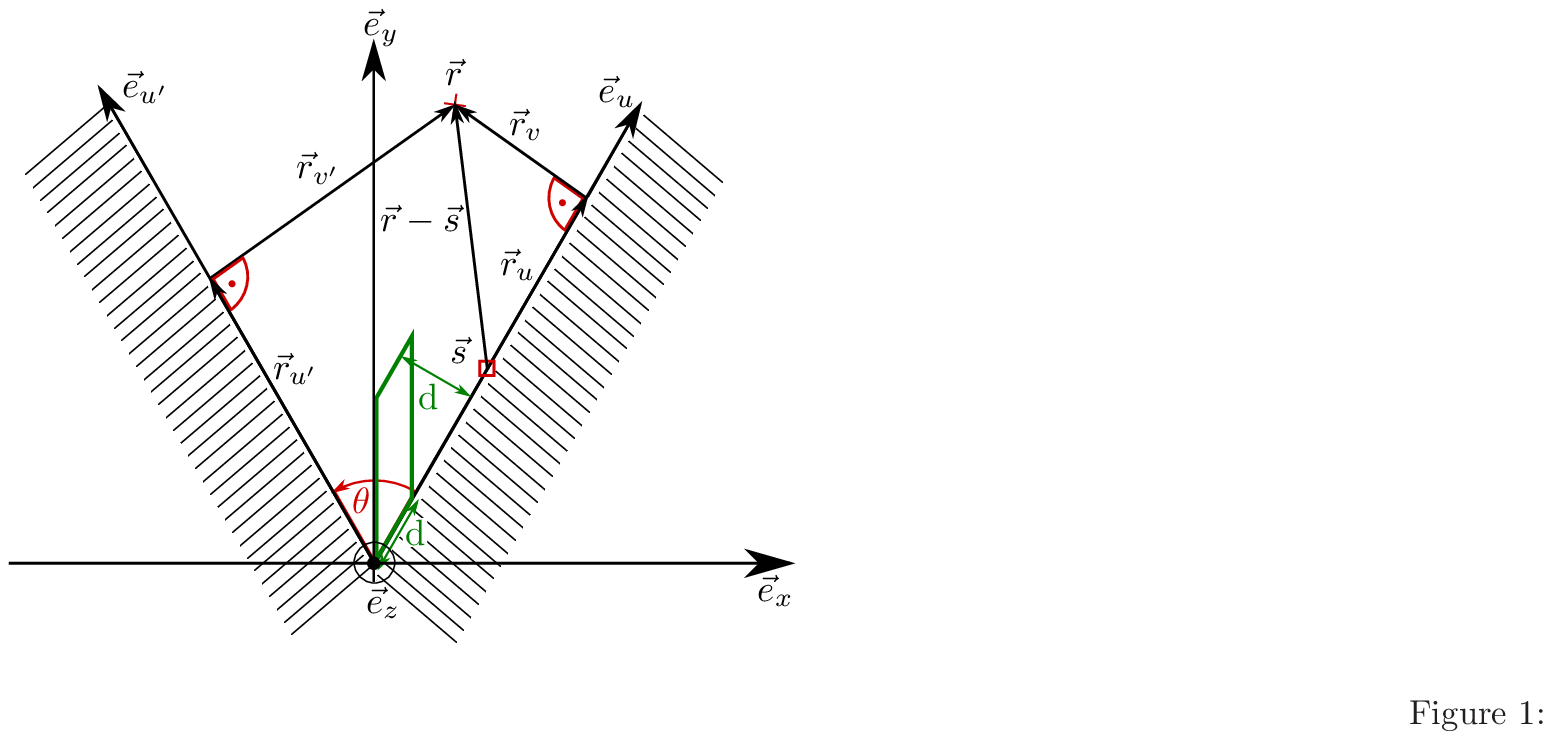}
	\caption{Schematic depiction of the studied system. 
		The two unit vectors $\vec{e}_u$ and $\vec{e}_{u'}$ are parallel to the two
		walls which meet at the opening angle $\theta$. 
		An arbitrary location $\vec{r}$ can be specified by the lateral and the normal
		components $\vec{r}_u,\vec{r}_v$ or $\vec{r}_{u'},\vec{r}_{v'}$ with respect to
		the walls.
		The parallelogram close to the wedge apex indicates the geometry of the unit
		cells by which the space in between the walls is tiled.
	\label{fig:ModelWedge}}
\end{figure}

In the present work, the filling behavior of an electrolyte solution close to a
wedge-like substrate is studied.
Consider in three-dimensional Euclidean space a wedge composed of two 
semi-infinite planar walls meeting at an opening angle $\theta$ along the
$z$-axis of a Cartesian coordinate system (see Fig.~\ref{fig:ModelWedge}).
Due to the translational symmetry in $z$-direction the system can be treated as
quasi-two-dimensional. 
In between the two walls an electrolyte solution composed of an uncharged
solvent (index ``0''), univalent cations (index ``+''), and univalent anions
(index ``-'') is present.
The wedge is in contact with a gas bulk at thermodynamic coexistence between
liquid and gas phase.
This choice of the thermodynamic parameters allows for two different filling
states of the wedge. 
From macroscopic considerations \cite{pomeau1986wetting, hauge1992macroscopic},
a critical opening angle
\begin{equation}
	\theta_C = \pi - 2\vartheta
	\label{eq:CriticalOpeningAngles}
\end{equation}
with the contact angle $\vartheta$ of the liquid can be derived, which
marks the transition between the wedge being filled by gas (``empty wedge'')
for $\theta > \theta_C$ and the wedge being filled by liquid for $\theta < 
\theta_C$. 
It is of utmost importance for the following to realize that, from the
microscopic point of view, a macroscopically empty wedge is typically partially filled by liquid.

\FloatBarrier
Characterizing the dependence of the critical opening angle $\theta_C$ on the
wall charge and describing the partial filling upon approaching the filling
transition for $\theta \gtrsim \theta_C$ are the objectives of the present
study.

%-------------------------------------------------------------------------------

\subsection{Density functional theory}

In order to determine the equilibrium structure of the fluid in terms of the
density profiles of the three species, classical density functional theory 
\cite{EVANS1979} is used. 
As wetting phenomena typically require descriptions on several length scales,
computational advantage is gained by studying a lattice fluid model in the
spirit of Refs.~\cite{Ibagon2013,Ibagon2014,Ibagon2016}.
In order to account for the special geometry of the system at hand, the
standard lattice fluid model is adapted by using parallelograms as basic
elements of the grid, which is indicated by the parallelogram close 
to the apex of the wedge in Fig.~\ref{fig:ModelWedge}. 
The size of an elementary parallelogram, which can be occupied by at most
one particle of either species, is chosen such that, with d denoting
the particle diameter, the sides parallel to the wall are of length d and
they are a distance d apart from each other (see Fig.~\ref{fig:ModelWedge}).
Each cell is identified by a pair $(l,j)$ of integer indices where $l \geq 0$
denotes the distance from the wall and $j$ represents the location parallel 
to the walls (see Fig.~\ref{fig:ModelWedge}).
The approximative density functional of this model used in the present work
can be written as
\begin{widetext}
	\begin{align}	
		\beta\Omega[\phi] =
		&\rho_\text{max} \text{d}^3\sum_{l,j}
		\Big[\sum_{\alpha \in \{0,\pm\}}\phi_{\alpha;l,j}
			(\ln(\phi_{\alpha;l,j})-\mu_\alpha^*+\beta V_{l,j})
			+ (1-\phi_{\text{tot};l,j})\ln(1-\phi_{\text{tot};l,j})
			\nonumber\\
			&+ \frac{1}{2}\sum_{n,m}\beta U^*_{l,j;n,m}
		\phi_{\text{tot};l,j}\phi_{\text{tot};n,m}\Big] 
		+ \beta U_\text{el},
		\label{eq:DFT1}
	\end{align}
\end{widetext}
where $\phi_{\alpha;l,j} = \rho_{\alpha;l,j} \text{d}^3$ denotes the packing fraction of fluid component
$\alpha\in\{0,\pm\}$ inside the cell specified by the indices $(l,j)$, $\phi_\text{tot} = \phi_0 + \phi_+ + \phi_-$
being the sum of the partial packing fractions, $\mu_\alpha^*$ is the effective chemical
potential of component $\alpha$, and $\rho_\text{max} = 1/\text{d}^3$ is the maximal 
number density of the fluid.
In the following the values $k_B T = 1/\beta$ with $T = \SI{300}{K}$ and $\rho_\text{max} = \SI{55.5}{\mol}$ are
chosen in correspondence with water at room temperature. 
Whereas the first line of Eq.~\eqref{eq:DFT1} corresponds to the exact lattice
fluid of non-interacting particles in an external field, the terms in the
second line of Eq.~\eqref{eq:DFT1} describe interactions amongst the particles
in a mean-field-like fashion.

The external potential $V_{l,j}$ in Eq.~\eqref{eq:DFT1} describes the
non-electrostatic interaction of the wall with a particle in cell $(l,j)$.
It is chosen to be independent of the specific particle type.
Here the wall-fluid interaction strength at a given position $\vec{r}$ results
from a superposition of interactions with all points $\vec{s}$ at the surface
of the walls (see Fig.~\ref{fig:ModelWedge}):
\begin{align}
	\beta V(\vec{r}) =
	&\int_0^\infty \mathrm{d}u  \beta \Phi(|\vec{r} - u \vec{e}_u   |) +
	\nonumber\\
	&\int_0^\infty \mathrm{d}u' \beta \Phi(|\vec{r} - u'\vec{e}_{u'}|),
	\label{eq:WALL1}
\end{align}
where $\beta\Phi$ is the underlying molecular pair potential of the wall-fluid 
interaction. 
For the sake of simplicity the Gaussian form
\begin{equation}
	\beta\Phi(r) \sim \exp\left(-\left(\frac{r}{\lambda}\right)^2\right)
	\label{eq:WallPot}
\end{equation}
with decay length $\lambda$ is used, which leads to the non-electrostatic 
wall-fluid interaction, 
Eq.~\eqref{eq:WALL1},
\begin{align}
	\beta V(\vec{r}) = h\Bigg(
	&\exp\left(-\left(\frac{r_v}{\lambda}\right)^2\right)
	\mathrm{erfc}\left(-\frac{r_u}{\lambda}\right) +
	\nonumber\\
	&\exp\left(-\left(\frac{r_{v'}}{\lambda}\right)^2\right)
	\mathrm{erfc}\left(-\frac{r_{u'}}{\lambda}\right) \bigg), 
	\label{eq:WALL5}
\end{align}
where the dimensionless coefficient $h$ describes the wall-fluid interaction strength.

The two remaining expressions in Eq.~\eqref{eq:DFT1} consider the interactions among the particles,
	which we consider as being composed of an electrically neutral molecular body and, in the case of the ions, an
	additional charge monopole.
	The way these interactions are treated regards the interactions as split in two contributions:
	the interaction between uncharged molecular bodies, which we refer to as non-electrostatic
	contribution, and the interaction between charge monopoles.
	In the present work we ignore the cross-interactions between a charge monopole and a neutral body.	
	However the chosen model proves to be sufficiently precise as it qualitatively captures the
	relevant feature of an increase of the ion density for an increasing solvent density.
	For example in the case of a liquid phase with density $\phi_0 = 0.80907$ coexisting with a gas phase with
	density $\phi_0 = 0.19093$, the ion densities increase from $\phi_\pm = 1.81541 \cdot 10^{-3}$ in the
gas to $\phi_\pm = 7.51554 \cdot 10^{-3}$ in the liquid.

In the Eq.~\eqref{eq:DFT1}, the non-electrostatic contribution to the 
fluid-fluid interaction is treated within random-phase approximation (RPA)
based on the interaction pair potential $U^*_{l,j;n,m}$ between a fluid
particle in cell $(l,j)$ and another one in cell $(n,m)$.
Here this interaction is assumed to be independent of the particle type and it
is assumed to act only between nearest neighbors, i.e., between particles located
in adjacent cells. 

Finally, in Eq.~\eqref{eq:DFT1} all electrostatic interactions, both wall-fluid
and fluid-fluid, are accounted for by the electric field energy $\beta U_\text{el}$.
The electric field entering $\beta U_\text{el}$ is determined by Neumann boundary
conditions set by a uniform surface charge density $\sigma$ at the walls of
the wedge, planar symmetry far away from the wedge symmetry plane and global
charge neutrality. Furthermore, the dielectric constant is assumed
	to be dependent on the solvent density. It is chosen to interpolate linearly
	between the values for vacuum ($\epsilon = 1$) and water ($\epsilon = 80$).
	This linear interpolation has been previously shown to match the behavior of 
	the dielectric constant in mixtures of fluids very well \cite{kaatze1984dielectric}. 
	In addition it is important to note, that here the surface charge is not
	caused by the dissociation of ionizable surface groups, i.e., charge regulation
	as in Ref.~\cite{behrens2001charge} is not relevant here, but it is
	assumed to be created by an external electrical potential, which is applied to the wall.
	One can imagine the wall being an electrode with the counter electrode being placed
far from the wall inside the fluid.

%-------------------------------------------------------------------------------

\subsection{Composition of the grand potential}

Upon minimizing the density functional $\beta\Omega[\phi]$ in 
Eq.~\eqref{eq:DFT1} one obtains the equilibrium packing fraction profiles 
$\phi^\text{eq}$, which lead to the equilibrium grand potential
$\beta\Omega^\text{eq}=\beta\Omega[\phi^\text{eq}]$ of the system.
This equilibrium grand potential can be decomposed into three contributions:
\begin{equation}
	\beta\Omega^\text{eq} = -\beta p V + \beta\gamma A + \beta\tau L.
	\label{eq:LineTensionDefinition}
\end{equation}
The first contribution $-pV$ with the pressure $p$ and the fluid volume $V$
equals the bulk energy contribution. 
It corresponds to the grand potential of an equally-sized system completely
filled with the uniform gas bulk state. 
The second term $\gamma A$ with the interfacial tension $\gamma$ and the total
wall area $A$ corresponds to the quasi-one-dimensional case of the gas being in
contact with a planar wall. 
The third contribution $\tau L$ with the line tension $\tau$ and the length $L$
of the wedge is the only contribution to the total grand potential, where the
influence of the wedge enters, and it is therefore of particular importance
in the present work.

\section{Results\label{sec:results}}

\subsection{Bulk phase diagram \label{sec:BulkPhaseDiagram}}

\begin{figure}[t]
	\includegraphics[width=8cm]{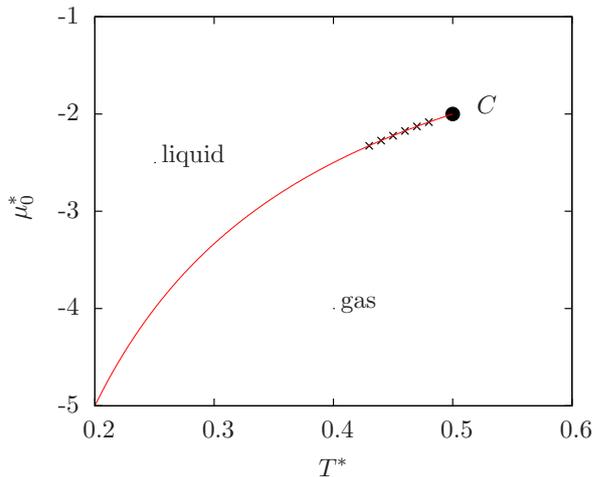}
	\caption{Bulk phase diagram in terms of the solvent chemical potential $\mu^*_0$ and the
		temperature $T^*$ for fixed ionic strength $I$.
		The solid red line represents the liquid-gas coexistence line for the salt-free 
		case ($I=0$), which is given by the analytical expression $\mu^*_0 = 
		-\frac{1}{T^*}$. 
		The black crosses indicate points of the liquid-gas coexistence curve for the
		case $I = \SI{5}{\milli\Molar}$. 
		The shift is small, which also holds for all ionic strengths used in this work 
		(up to $I=\SI{100}{\milli\Molar}$).
	\label{fig:BulkPhaseDiagram}}
\end{figure}

In the bulk region, far from any confinements, the densities $\phi_\alpha,
\alpha\in\{0,\pm\}$ of the three fluid components become constant, and, due
to local charge neutrality, $\phi_+ = \phi_-$. 
This simplifies the density functional $\beta\Omega[\phi]$ in 
Eq.~\eqref{eq:DFT1}, and the Euler-Lagrange equations read 
\begin{equation}
	\mu_\alpha^* = 
	\ln\frac{\phi_\alpha}{1 - \phi_\text{tot}} - \frac{2}{T^*}\phi_\text{tot},	
\end{equation}
where $1/T^*$ is proportional to the strength of the fluid-fluid interaction $\beta U^*$.
For the ion-free case $I = 0$ the liquid-gas coexistence line is given by
the analytical expression $\mu_0^* = -\frac{1}{T^*}$ (see solid red line in 
Fig.~\ref{fig:BulkPhaseDiagram}). 
For fixed but non-vanishing ionic strengths $I$ the liquid-gas coexistence
lines have been calculated numerically (see the black crosses in 
Fig.~\ref{fig:BulkPhaseDiagram}). 
Whereas the deviations from the ion-free case are only marginal in the bulk
phase diagram for all ionic strengths considered here, it is of major
importance to determine the coexistence conditions precisely, because surface
and line properties (see Eq.~\eqref{eq:LineTensionDefinition}) are highly
sensitive to them.

%-------------------------------------------------------------------------------

\begin{figure}[t]
	\includegraphics[width=8cm]{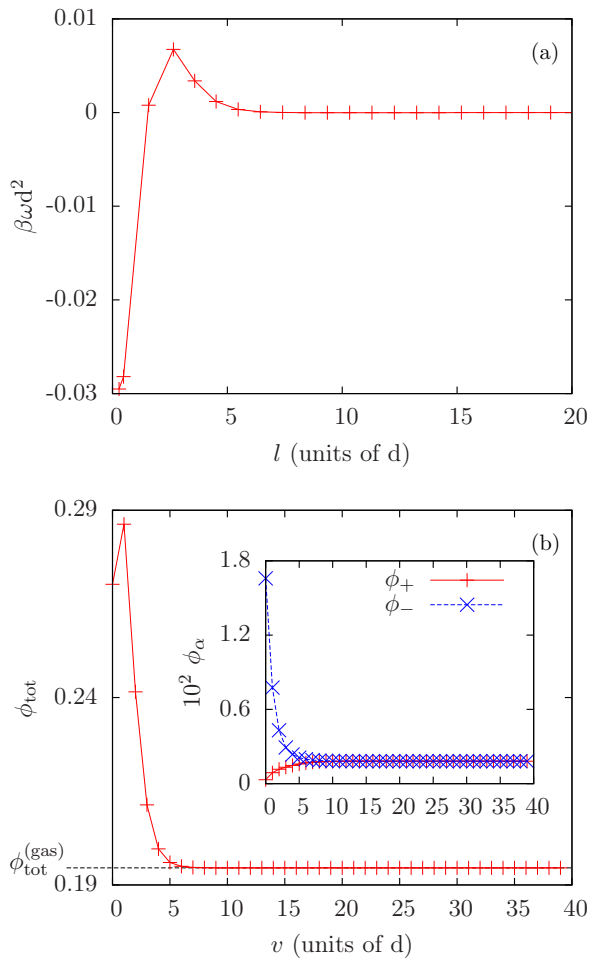}
	\caption{
		(a) Effective interface potential $\beta\omega$ as a function of the film 
		thickness $l$ of a liquid film between a planar wall and the gas bulk phase
		and (b) equilibrium total packing fraction profile $\phi_\text{tot}(v)$,
		with $v$ denoting the distance from the wall (see Fig. \ref{fig:ModelWedge}),
		corresponding to the minimum of $\beta\omega(l)$ for ionic strength $I = 
		\SI{100}{\milli\Molar}$, temperature $T^*=0.43$, wall-fluid interaction
		strength $h=0.09327$, decay length $\lambda = \SI{2}{\day}$, and surface charge
		density $\sigma = 
		\SI[per-mode=symbol]{0.03}{\elementarycharge\per\square\day}$. The inset in panel (b)
		shows the corresponding ion packing fractions $\phi_+$ and $\phi_-$ as functions of the distance $v$ from the wall.
		Panel (a) identifies the system exhibiting partial wetting for the present
	configuration.\label{fig:DensityAndOmega}}
\end{figure}

\subsection{Electrolyte wetting on a planar wall \label{sec:PlanarWall}}

Before studying the filling behavior of a wedge, it is important to study the
wetting of a planar wall because the results enter as the surface contributions
to the total grand potential Eq.~\eqref{eq:LineTensionDefinition} and the
quasi-one-dimensional packing fraction profiles provide the boundary conditions far
away from the wedge symmetry plane.
In the case of a planar wall the density functional $\beta\Omega[\phi]$ 
simplifies to a quasi-one-dimensional one and, due to the corresponding
relations $r_u=-r_{u'}, r_v=r_{v'}$ (see Fig.~\ref{fig:ModelWedge}), the expression Eq.~\eqref{eq:WALL5}
for the fluid-wall interaction becomes 
\begin{equation}
	\beta V(\vec{r}) = 2h\exp\left(-\left(\frac{r_v}{\lambda}\right)^2\right).
	\label{eq:betaV}
\end{equation}
With this set of equations one can determine the equilibrium packing fraction
$\phi_{\alpha;i}$ of the fluid close to the planar wall, where the integer index $i \geq 0$ denotes
the distance of the cell from the wall. 

One possibility to characterize wetting of a planar wall is by means of the
excess adsorption
\begin{equation}
	\Gamma[\phi_\text{tot}] \coloneqq 
	\sum_{i=0}^\infty(\phi_{\text{tot};i} - \phi_\text{tot}^\text{(gas)})
	\label{eq:ExcessDefinition} 
\end{equation}
with the total packing fraction $\phi_\text{tot}^\text{(gas)}$ of the gas phase
at liquid-gas coexistence for the given temperature $T^*$, which measures the
additional amount of particles in excess to the gas bulk phase due to the 
presence of the wall.
Alternatively, one can consider the film thickness
\begin{equation}
	l[\phi_\text{tot}] := 
	\frac{\Gamma[\phi_{\text{tot}}]}{\phi_\text{tot}^\text{(liquid)} -
	\phi_\text{tot}^\text{(gas)}}
	\label{eq:FilmThicknessDefinition}
\end{equation}
with the total packing fraction $\phi_\text{tot}^\text{(liquid)}$ of the liquid
phase at liquid-gas coexistence for the given temperature $T^*$, which 
corresponds to the thickness of a uniform liquid film of packing fraction
$\phi_\text{tot}^\text{(liquid)}$ with the same excess adsorption 
$\Gamma[\phi_\text{tot}]$ as the equilibrium total packing fraction profile 
$\phi_\text{tot}$.

Minimizing the grand potential functional Eq.~\eqref{eq:DFT1} for a planar
wall (see Eq.~\eqref{eq:betaV}) with the constraint of fixed excess adsorption
$\Gamma[\phi_\text{tot}]$, Eq.~\eqref{eq:ExcessDefinition}, or fixed film thickness 
$l[\phi_\text{tot}]$, Eq.~\eqref{eq:FilmThicknessDefinition}, and subtracting
the bulk contribution of the grand potential as well as 
the wall-liquid and the liquid-gas interfacial tensions ($\gamma_{sl}$ and $\gamma_{lg}$, respectively), one
obtains the effective interface potential $\beta\omega$ \cite{Dietrich1988}.
An example for $\beta\omega(l)$ is displayed in Fig.~\ref{fig:DensityAndOmega}(a). 
The position $l=l_{\text{eq}}$ of the minimum of the effective interface 
potential $\beta\omega(l)$ corresponds to the equilibrium film thickness.
The corresponding equilibrium total packing fraction profile $\phi_\text{tot}$
for the parameters chosen in Fig.~\ref{fig:DensityAndOmega}(a) is shown in 
Fig.~\ref{fig:DensityAndOmega}(b). 

\begin{figure}[t]
	\includegraphics[width=8cm]{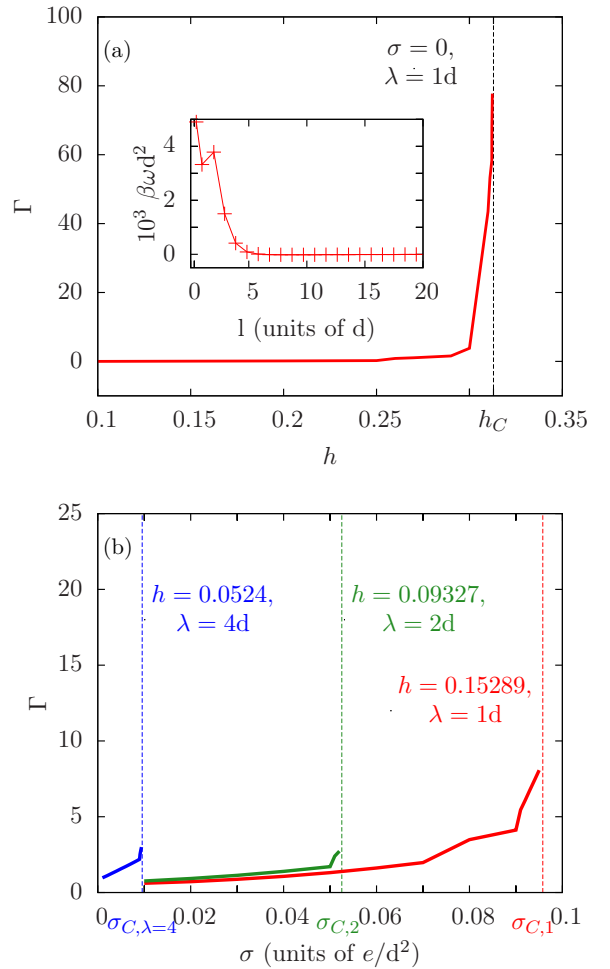}
	\caption{Excess adsorption $\Gamma$ (see Eq.~\eqref{eq:ExcessDefinition}) for (a) an
		electrically neutral planar wall ($\sigma=0$) and decay length $\lambda=\SI{1}{\day}$ as function
		of the wall-fluid interaction strength $h$ and (b) three different sets of  
		wall-fluid interaction strength $h$ and decay length $\lambda$ as function of
		the wall charge density $\sigma$. 
		Both panels exhibit an increase of the excess adsorption $\Gamma$ upon 
		approaching critical values $h_C$ or $\sigma_C$, respectively, at which the 
		system undergoes a wetting transition. 
		The discontinuity of $\Gamma$ at the critical values identifies the wetting
		transition to be of first order. This can also be verified by considering the 
		effective interface potential $\beta\omega(l)$, as shown in the inset in panel (a) for conditions slightly
		above the wetting transition.
		The barrier separating the local minimum at small film thickness l from the global minimum at
	large film thickness l proves the first order nature of the wetting transition.
	\label{fig:GammaSigma}}
\end{figure}

\begin{figure}[t]
	\includegraphics[width=8cm]{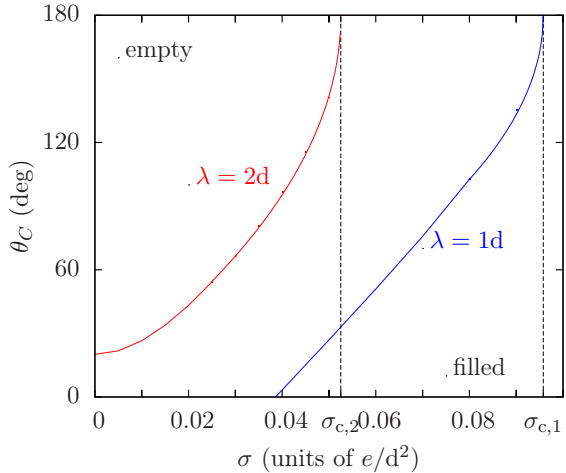}
	\caption{Critical opening angle $\theta_C$ of the wedge, at which the filling transition
		occurs, as function of the wall charge density $\sigma$ for decay lengths
		$\lambda \in \{\SI{1}{\day},\SI{2}{\day}\}$. 
		For $\theta > \theta_C$ the wedge is macroscopically empty, whereas for $\theta < \theta_C$ it is filled by liquid.
		The values $\theta_C$, derived via Eqs.~\eqref{eq:CriticalOpeningAngles} and
		\eqref{eq:contactangle}, increase with increasing wall charge $\sigma$. 
		At wall charge density $\sigma=\sigma_C$ the critical angle of the filling
		transition is $\theta_C = \ang{180}$, i.e., the filling transition is actually
		the wetting transition of a planar wall (see Sec.~\ref{sec:PlanarWall}).
	\label{fig:CriticalOpeningAngle}}
\end{figure}

Using this procedure, one can determine the equilibrium density profiles for
different ionic strengths $I$, temperatures $T^*$, wall-fluid interaction 
strengths $h$, decay lengths $\lambda$, and surface charge densities $\sigma$. 
Figure~\ref{fig:GammaSigma}(a) displays the equilibrium excess adsorption 
$\Gamma$ as function of the wall-fluid interaction strength $h$ for surface
charge density $\sigma=0$ and decay length $\lambda=\SI{1}{\day}$. 
Due to the vanishing surface charge, the packing fraction profiles of the two
ion species are identical, $\phi_+=\phi_-$, hence the fluid is locally charge
neutral and the electrostatic energy $\beta U_\text{el}$ in Eq.~\eqref{eq:DFT1}
vanishes.
Therefore, due to the small number densities of the ions, this case is similar
to an ion-free system, where a wetting transition is caused by an increase of
the non-electrostatic wall-fluid interaction strength $h$ (see 
Eq.~\eqref{eq:betaV}) up to a critical value $h_C$. 
In contrast, Fig.~\ref{fig:GammaSigma}(b) shows the excess adsorption 
$\Gamma$ for different sets of the wall-fluid interaction strength $h$ and the
decay length $\lambda$ as function of the surface charge density $\sigma$.
The values of $h$ are chosen in such a way, that the three respective decay
lengths $\lambda = $\SIlist{1;2;4}{\day} lead to the same values of the volume
integrals of the corresponding wall-fluid interaction potentials,
\begin{equation}
	\int_{\mathcal{V}}\text{d}r\ \beta V(\vec{r}).
\end{equation}
Here, the wall charge $\sigma$ is varied and a wetting transition is observed
at a critical value $\sigma_C$. 

\begin{figure*}[t]
	\includegraphics[width=16cm]{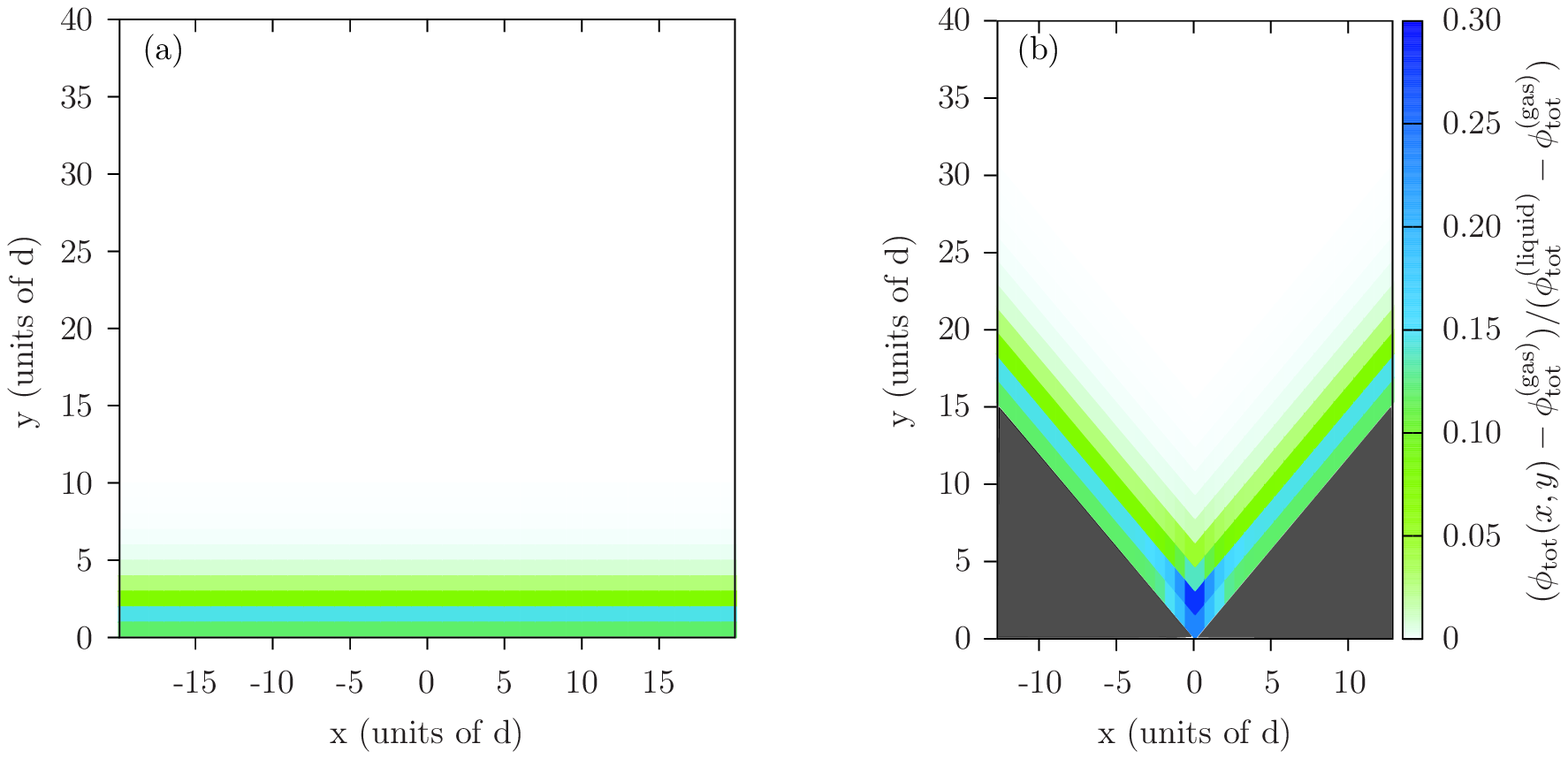}
	\caption{Equilibrium packing fraction profiles $\phi_{\text{tot}}(x,y)$ inside a wedge with opening
		angle (a) $\theta = \ang{180}$ and (b) $\theta = \ang{80}$ for
		ionic strength $I = \SI{100}{\milli\Molar}$, wall-fluid interaction
		strength $h = \SI{0.09327}{}$, decay length $\lambda = 
		\SI{2}{\day}$, and wall charge density $\sigma = 
		\SI[per-mode=symbol]{0.03}{\elementarycharge\per\square\day}$. 
		Far away from the symmetry plane of the wedge the packing fraction
		profiles coincide with those at planar walls (see 
		Fig.~\ref{fig:DensityAndOmega}(b)).
		Upon decreasing the opening angle $\theta$, an increase of the
	density close to the tip of the wedge occurs (see panel (b)).\label{fig:DensityProfiles}}
\end{figure*}

All four setups in Fig.~\ref{fig:GammaSigma} exhibit the characteristics of first-order wetting
transitions, which are identified by finite limits of $\Gamma$ upon 
$h\nearrow h_C$ or $\sigma\nearrow\sigma_C$. In addition for all these cases the first-order nature
has been verified by studying the effective interface potential (see inset in Fig.~\ref{fig:GammaSigma}(a)),
which is clearly manifested by the energy barrier separating the local and the global minimum.
For the quasi-ion-free case $\sigma=0$ in Fig.~\ref{fig:GammaSigma}(a) the
choice Eq.~\eqref{eq:WallPot} of the molecular pair potential of the wall-fluid 
interaction leads to a wetting transition of first order, in contrast to the
choice of the nearest neighbor potential in Ref.~\cite{Ibagon2013}, which
generates a second-order wetting transition.
However, it has been shown that for $\sigma\neq0$ (see 
Fig.~\ref{fig:GammaSigma}(b)) wetting transitions are of first order
once the Debye length is larger than the bulk correlation length 
\cite{Ibagon2014}.

%-------------------------------------------------------------------------------

\subsection{Wedge wetting by an electrolyte solution \label{sec:WedgeWetting}}

Having studied the system under consideration in the bulk 
(Sec.~\ref{sec:BulkPhaseDiagram}) and close to a planar wall 
(Sec.~\ref{sec:PlanarWall}), one can investigate wedge-shape geometries.
As explained in the context of Eq.~\eqref{eq:CriticalOpeningAngles}, the
system undergoes a filling transition for the  opening angle $\theta$
(see Fig.~\ref{fig:ModelWedge}) approaching the critical opening angle 
$\theta_C$ from above.
For $\theta<\theta_C$ the wedge is macroscopically filled by liquid, whereas
for $\theta>\theta_C$ the wedge is macroscopically empty.
In the following, the filling of an empty wedge, i.e., $\theta\searrow\theta_C$,
will be studied.

Following Eq.~\eqref{eq:CriticalOpeningAngles}, the critical opening angle
$\theta_C$  can be calculated from the contact angle $\vartheta$ of the 
liquid, which is related to the depth of the minimum of the effective interface
potential by \cite{Dietrich1988}
\begin{align}
	\cos\vartheta = 1 + \frac{\omega(l_\text{eq})}{\gamma_{lg}}
	\label{eq:contactangle}
\end{align}
with the liquid-gas surface tension $\gamma_{lg}$.
Hence, the critical opening angle $\theta_C$ can be inferred from the wetting
properties of a planar wall using the method of Sec.~\ref{sec:PlanarWall}.
Figure~\ref{fig:CriticalOpeningAngle} displays the critical opening angle
$\theta_C$ as function of the wall charge $\sigma$ for the case of decay 
lengths $\lambda \in \{\SI{1}{\day},\SI{2}{\day}\}$.
As the contact angle $\vartheta$ decreases upon increasing the wall charge due
to the electrowetting effect \cite{Bier2014}, the critical opening angle 
$\theta_C$ increases with increasing wall charge. 
For the critical wall charge $\sigma=\sigma_C$ the critical opening angle 
$\theta_C$ reaches the value of \ang{180}, since for this wall charge the
wetting transition of the planar wall occurs (compare 
Fig.~\ref{fig:GammaSigma}(b)), i.e., for a planar wall the wetting and
the filling transition are identical.

Figure~\ref{fig:DensityProfiles} displays the equilibrium packing fraction profiles inside wedges
with opening angles $\theta = \ang{180}$ (Fig.~\ref{fig:DensityProfiles}(a)) and $\theta = 
\ang{80}$ (Fig.~\ref{fig:DensityProfiles}(b)) with the parameters $h$, $\lambda$, and 
$\sigma$ identical to those of Fig.~\ref{fig:DensityAndOmega}(b).
Away from the wedge symmetry plane the structure rapidly converges towards
that of a planar wall, which verifies the chosen size of the numerical grid
being sufficiently large to capture all interesting effects.
Furthermore, the decrease of the opening angle, as shown in 
Fig.~\ref{fig:DensityProfiles}(b), leads to an increase of the density close to the tip of
the wedge. 
For example the maximal density increases from \SI{15}{\percent} of the relative density
difference between liquid and gas density to almost \SI{30}{\percent}. 
However, the increase in the density is limited to the close vicinity of the
tip of the wedge, which is an indication of first-order filling transitions.
In fact, in the presence of ions, wetting transitions at a planar wall are of
first order with a barrier in the effective interface potential $\beta \omega(l)$ (see Fig.~\ref{fig:DensityAndOmega}(a))
being present for all states below the wetting transition of a planar wall \cite{Ibagon2014}.
Hence filling transitions of wedges are expected
to be of first order, too \cite{parry2000critical, parry2001interfacial}.  

\begin{figure}[t]
	\includegraphics[width=8cm]{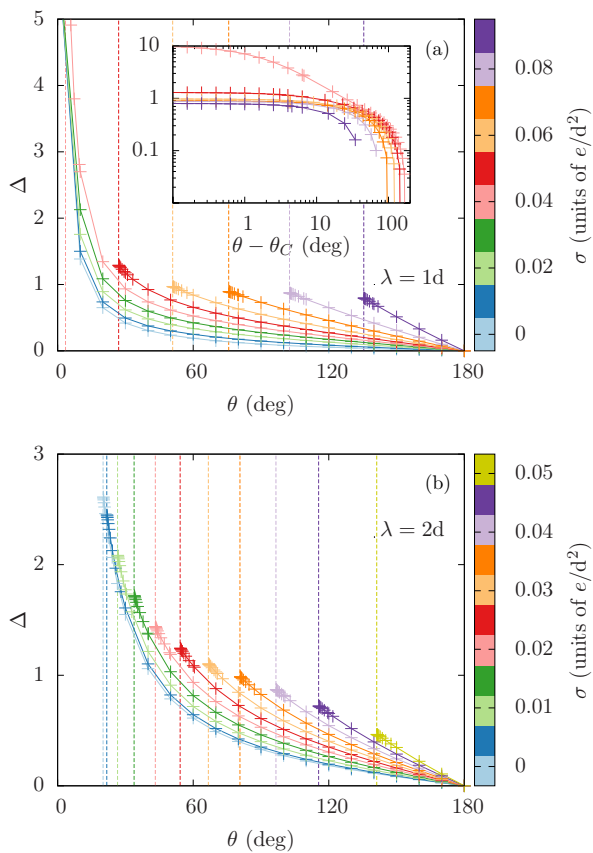}
	\caption{Wedge adsorption $\Delta$ (see Eq.~\eqref{eq:DeltaDefinition}) as function of
		the opening angle $\theta$ of the wedge and of the wall charge density 
		$\sigma$. 
		In panel (a) the decay length of the wall-fluid interaction potential is 
		$\lambda = \SI{1}{\day}$, whereas in panel (b) it is $\lambda = \SI{2}{\day}$.
		Similar to the filling height $l_w$ shown in Fig.~\ref{fig:FillingHeights}, the wedge adsorption 
		$\Delta$ increases for increasing wall charge density $\sigma$ as well as for
		decreasing opening angle $\theta$.
		The limits of $\Delta$ upon approaching the filling transition, $\theta
		\searrow\theta_C$, are finite, which signals a first-order filling 
		transition. To highlight this, the inset in panel (a) shows a double-logarithmic
	plot of the wedge adsorption $\Delta$ as function of the distance $\theta - \theta_C$ from the filling transition.\label{fig:Deltas}}
\end{figure}

In order to describe the filling transition of a wedge quantitatively, several
quantities have been studied. 
Firstly the wedge adsorption 
\begin{equation}
	\Delta = 
	\sum_i \sum_j(\phi_{i,j} - \phi_{\text{tot}}^{\text{(gas)}}) - 
	\Gamma l_\text{wall} / d,
	\label{eq:DeltaDefinition}
\end{equation}
with the length of the wall $l_{\text{wall}}$ shall be discussed. 
In the spirit of the excess adsorption $\Gamma$ at a planar wall 
(Eq.~\eqref{eq:ExcessDefinition}), this quantity $\Delta$ measures the excess
of an inclined wedge above the excess adsorption $\Gamma$ of a planar wall. 
In Fig.~\ref{fig:Deltas} the wedge adsorption $\Delta$ is shown as function of the opening
angle $\theta$ and of the wall charge density $\sigma$ for decay lengths 
$\lambda=\SI{1}{\day}$ (Fig.~\ref{fig:Deltas}(a)) and $\lambda=\SI{2}{\day}$ 
(Fig.~\ref{fig:Deltas}(b)). 
The ionic strength is $I=\SI{100}{\milli\Molar}$ and the wall-fluid interaction strength $h$ has
been chosen as in Fig.~\ref{fig:GammaSigma}(b).
Upon decreasing the opening angle $\theta$ the wedge adsorption $\Delta$
increases, regardless of the wall charge density $\sigma$, the decay length
$\lambda$, or the non-electrostatic wall-fluid interaction strength $h$.
However, the limits of $\Delta$ upon approaching the filling transition, 
$\theta\searrow\theta_C$, are finite, which signals a first-order filling 
transition (see in particular the inset of Fig.~\ref{fig:Deltas}(a)).
Moreover, for any fixed opening angle $\theta>\theta_C$, the wedge adsorption
$\Delta$ increases with increasing wall charge density $\sigma$. 
Both observations can be understood in terms of the strength of the interaction
between wall and fluid. 
In case of an increasing wall charge density $\sigma$, the increase of $\Delta$
stems from an increase of the counterion density which is stronger than 
the accompanying decrease of the coion density.
This phenomenon is well-known for non-linear Poisson-Boltzmann-like theories as
the present one.
For the case of a decreasing opening angle $\theta>\theta_C$ the growing
overlap of the wall-fluid interactions, both the non-electrostatic as well as
the electrostatic one, leads to an increase in the density.

Besides these general qualitative trends there are quantitative differences
for the two cases in Fig.~\ref{fig:Deltas}, which differ in the values of the decay length 
$\lambda$.
One way to compare Figs.~\ref{fig:Deltas}(a) and \ref{fig:Deltas}(b) is to
consider the limits $\Delta(\theta_C^+)$ upon $\theta\searrow\theta_C$ for a
common value of the wall charge density $\sigma$. 
In this case, the shorter-ranged wall-fluid interaction, $\lambda = 
\SI{1}{\day}$ (see Fig.~\ref{fig:Deltas}(a)), leads to higher values of 
$\Delta(\theta_C^+)$ than the longer-ranged one, $\lambda = \SI{2}{\day}$ (see
Fig.~\ref{fig:Deltas}(b)).
However, since shorter decay lengths $\lambda$ lead to smaller critical opening
angles $\theta_C$ (see Fig. \ref{fig:CriticalOpeningAngle}), which correspond to stronger overlaps of the wall-fluid
interactions of the two walls of the wedge, an increase in the wedge adsorption
$\Delta$ is caused mostly for geometrical reasons. 
Alternatively, if one compares Fig.~\ref{fig:Deltas}(a) and \ref{fig:Deltas}(b)
for a fixed opening angle $\theta>\theta_C$ and a fixed wall charge density
$\sigma$, the wedge adsorption $\Delta$ is larger for the case of the 
longer-ranged wall-fluid interaction.
This can be readily understood given the fact that, for fixed opening angle 
and wall charge, the interaction strength at a specific point in the system is
the stronger the longer ranged the interaction is. 

\begin{figure}[t]
	\includegraphics[width=8cm]{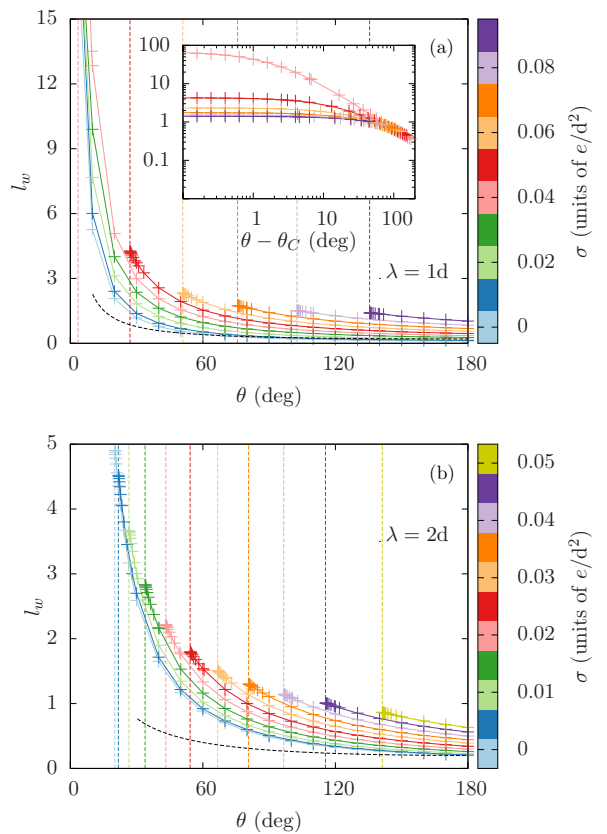}
	\caption{Filling height at the symmetry plane $l_w$ as function of the opening angle
		$\theta$ of the wedge and of the wall charge density $\sigma$. 
		In panel (a) the decay length of the wall-fluid interaction is $\lambda = 
		\SI{1}{\day}$, whereas it is $\lambda = \SI{2}{\day}$ in panel (b). 
		The dashed black curve in both panels corresponds to the thickness of the first
		layer of cells on the symmetry axis. 
		The comparison of this curve with the curves of the filling height $l_w$
		shows, that the increase of $l_w$ close to the critical opening angle
		$\theta\gtrsim \theta_C$ stems from the increasing interactions close to the
		tip of the wedge.
		Furthermore the filling height $l_w$ increases with both an increasing wall
		charge density $\sigma$ as well as a decreasing opening angle $\theta$.
		The finite limits for $l_w$ upon $\theta\searrow\theta_C$ point to a 
		first-order filling transition. Similarly in Fig.~\ref{fig:Deltas}(a) the inset in panel (a)
	shows a double-logarithmic plot of the filling height $l_w$ as function of the distance $\theta - \theta_C$ 
	from the filling transition to verify its first-oder nature.\label{fig:FillingHeights}}
\end{figure}

As a second quantity to describe the filling of a wedge the filling height
\begin{equation}
	l_w = 
	\frac{\Gamma_\text{sym}}{\phi_\text{tot}^\text{(liquid)} - 
	\phi_\text{tot}^\text{(gas)}}
\end{equation}
is considered, where $\Gamma_\text{sym}$ denotes the excess adsorption along
the symmetry plane (cell index $j=0$) of the wedge:
\begin{align}
	\Gamma_\text{sym} := 
	\sum_{l}(\phi_{\text{tot};l,0} - \phi_\text{tot}^\text{(gas)}).
\end{align}
The definition of the filling height $l_w$ of a wedge is similar to that of the film
thickness $l$ at a planar wall (see Eq.~\eqref{eq:FilmThicknessDefinition}).
It expresses the distance of the liquid-gas interface of the adsorbed film
from the tip of the wedge.
Figure~\ref{fig:FillingHeights} displays the filling height $l_w$ as function of the opening angle
$\theta$ and of the wall charge $\sigma$ with the decay lengths
$\lambda=\SI{1}{\day}$ in Fig.~\ref{fig:FillingHeights}(a) and $\lambda=
\SI{2}{\day}$ in Fig.~\ref{fig:FillingHeights}(b). 
When discussing the filling height $l_w$ one has to account for the geometrical
effect of an increasing side length $l_{w1}(\theta):=d/\sin(\theta/2)$ of the
elementary parallelograms in the direction of the symmetry plane (see Fig.~\ref{fig:ModelWedge})
upon decreasing the opening angle $\theta$.
It is equivalent to a filling height of exactly one cell and it is displayed
in Fig.~\ref{fig:FillingHeights} as a black dashed curve.
By comparing the filling height $l_w(\theta)$ with the trend given by the side
length $l_{w1}(\theta)$ one infers a stronger increase of the former upon
approaching the filling transition $\theta\searrow\theta_C$, which can be
attributed to the filling effect.
Similar to the wedge adsorption $\Delta$, the filling height $l_w$ increases 
either upon decreasing the opening angle $\theta$ towards the critical opening
angle $\theta_C$ or, for fixed $\theta>\theta_C$, upon increasing the magnitude
of the wall charge density $\sigma$.
The reason for these observed trends of the filling height $l_w$ is again, as
for the wedge adsorption $\Delta$, a consequence of the increased magnitude of 
the wall-fluid interaction.
Finally, the filling height $l_w$, as the wedge adsorption, approaches a finite
limit upon $\theta\searrow\theta_C$, which is in agreement with the expectation
of a first-order filling transition.

\begin{figure}[t]
	\includegraphics[width=8cm]{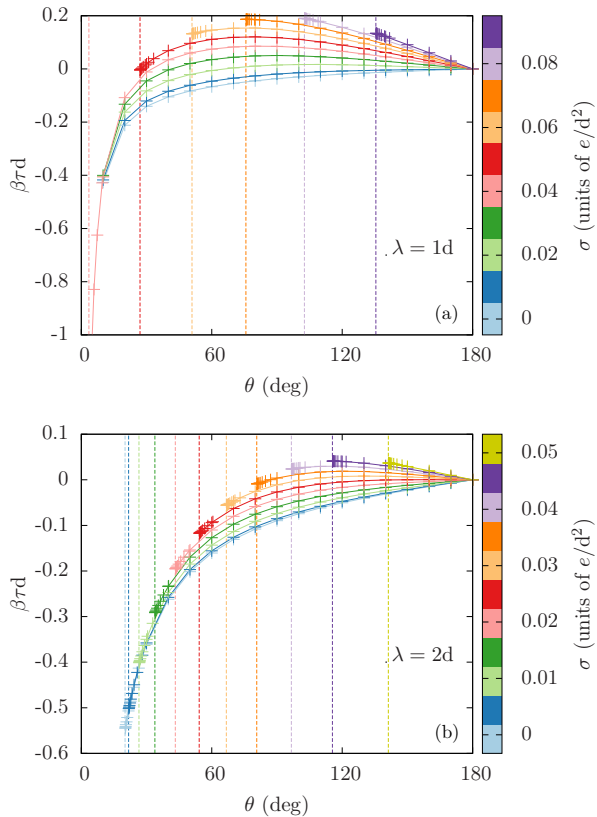}
	\caption{Line tension $\tau$ as function of the opening angle $\theta$ and of the wall
		charge density $\sigma$ for decay lengths (a) $\lambda = \SI{1}{\day}$ and 
		(b) $\lambda = \SI{2}{\day}$. 
		For small wall charge density $\sigma$, the line tension $\tau$ is negative for
		all opening angles $\theta$, whereas for sufficiently large $\theta$ and 
	$\sigma$ positive values of $\tau$ may occur.\label{fig:LineTensions}}
\end{figure}
% \FloatBarrier

As shown in Eq.~\eqref{eq:LineTensionDefinition}, the equilibrium grand 
potential $\Omega^\text{eq}$ may contain a contribution scaling proportional
to a linear extension $L$ of the system and the corresponding coefficient of
proportionality of the dimension of an energy per length is called the line
tension $\tau$.
In the present context of a wedge, the line tension $\tau$ measures the 
structural difference between a wedge and a planar wall, and the contribution
$\tau L$ scales with the length $L$ of the wedge along the $z$-direction.

Figure~\ref{fig:LineTensions} displays the line tension $\tau$ as function of the opening angle
$\theta$ and of the wall charge density $\sigma$ for decay lengths $\lambda =
\SI{1}{\day}$ (Fig.~\ref{fig:LineTensions}(a)) and $\lambda = \SI{2}{\day}$ 
(Fig.~\ref{fig:LineTensions}(b)). 
The qualitative dependence of the line tension $\tau$ on the opening angle
$\theta$ turns out to depend on the wall charge density $\sigma$:
For small wall charge densities the line tension is negative and it decreases
monotonically with decreasing opening angle.
For sufficiently large wall charge densities the line tension is positive for
large opening angles and, if the critical opening angle $\theta_C$ is small
enough, negative for small opening angles, i.e., the line tension may depend
non-monotonically on the opening angle.
For molecular length scales $d\approx3\,\text{\AA}$ and room temperature 
$T\approx300\,\text{K}$ the order of magnitude of the line tension 
$|\tau|\approx\si{\pico\newton}$ is in accordance with literature 
\cite{Ibagon2016,getta1998line,dussaud1997wetting}. 

\section{Conclusions and summary\label{sec:conclusion}}

In the present work the filling of charged wedges by electrolyte
solutions has been studied within microscopic classical density functional
theory of a lattice model (Fig.~\ref{fig:ModelWedge}).
As in previous studies \cite{Ibagon2013, Ibagon2014, Ibagon2016}, considering
lattice models offers technical advantages over continuum models, as the former
allow for the explicit description of larger parts of the system.
The electrolyte solution comprises a solvent and a univalent salt.
A short-ranged attractive interaction between the fluid particles leads to a
liquid-gas phase transition of the bulk electrolyte solution (Fig.~\ref{fig:BulkPhaseDiagram}).
A fluid-wall interaction derived from a Gaussian pair potential (Eq.~\eqref{eq:WallPot}) gives
rise to first-order wetting transitions of a planar wall in contact with a
gas bulk phase (Figs.~\ref{fig:DensityAndOmega}).
This first-order wetting transition of a planar wall can be driven by the
wall-fluid interaction strength or by the surface charge density (Fig.~\ref{fig:GammaSigma}).
The critical opening angle, below which the wedge is filled, depends on the
surface charge density and on the decay length of the wall-fluid interaction
(Fig.~\ref{fig:CriticalOpeningAngle}).
Upon approaching the critical opening angle from above, a macroscopically small
but microscopically finite amount of fluid is accumulated close to the apex of
the wedge (Fig.~\ref{fig:DensityProfiles}).
This observation as well as the finite limits of the wedge adsorption (Fig.~\ref{fig:Deltas}),
the filling height (Fig.~\ref{fig:FillingHeights}), and the line tension (Fig.~\ref{fig:LineTensions}) are compatible with
a first-order filling transition.
Upon increasing the surface charge density, the line tension as function of the
opening angle changes from a monotonically increasing negative function via
a function exhibiting a positive maximum to a monotonically decreasing positive
function (Fig.~\ref{fig:LineTensions}).

The unequivocally first-order filling transitions found within the model of the
present work are in full agreement with the general expectation for systems with
barriers in the effective interface potential at the filling transition
\cite{parry2000critical, parry2001interfacial}.
Moreover, this is expected to be the case for any electrolyte solution not too
close to a critical point, as such systems exhibit barriers in the effective
interface potential for all conditions of partial wetting \cite{Ibagon2014}.  
Therefore, the optimistic point of view in Ref.~\cite{malijevsky2013critical}
expecting the experimental accessibility of systems displaying critical filling 
transitions requires to exclude the vast class of dilute electrolyte solutions
as potential candidates.
On the other hand, being assured of the first-order nature of filling 
transitions in the presence of electrolyte solutions allows one to numerically
efficiently set up more realistic models, which are not restricted to a lattice
for technical reasons, to quantitatively describe wetting and filling of 
complex geometries. 

\bibliography{Bibliography}

\end{document}